\documentclass[a4paper,pra,reprint,notitlepage,superscriptaddress, nofootinbib, showpacs, eqsecnum]{revtex4-1}
\usepackage[stable]{footmisc}
\usepackage{array}
\usepackage{verbatim}
\usepackage{tikz}
\usepackage{hyperref}
\usepackage{color}
\usepackage{multirow}
\usepackage{graphics}
\usepackage{epsfig}
\usepackage{amsmath} % AMS Math Package
\usepackage{amsthm} % Theorem Formatting
\usepackage{amssymb}	% Math symbols such as \mathbb

\usepackage{graphicx}% Include figure files
\usepackage{dcolumn}% Align table columns on decimal point

\usepackage{graphicx}
\usepackage{epsfig}
\usepackage{wrapfig}
\usepackage{mathrsfs}

\newcommand{\bla}{\color{black}}

\newcommand{\defcol}{\bla}%{\color{blue!80!black}} 
\newcommand{\ampcol}{\bla}%{\color{red!80!black}} 
\newcommand{\concol}{\bla}%{\red}
\newcommand{\errcol}{\bla}%{\bla}  

\newcommand{\Htot}{ {H}}
\newcommand{\Hc}{\concol {H}_c\bla}
\newcommand{\Herr}{\errcol {H}_0\bla}

\newcommand{\Ba}{\ampcol\beta_{_{\Omega}}\bla}
\newcommand{\Bd}{\defcol\beta_z\bla}
\newcommand{\Sa}{\ampcol S_{_{\Omega}}\bla}
\newcommand{\Sd}{\defcol S_z\bla}
\newcommand{\Fa}{\ampcol F_{_{\Omega}}\bla}
\newcommand{\Fd}{\defcol F_z\bla}

\newcommand{\ket}[1]{\left| #1 \right>} % for Dirac bras
\newcommand{\FT}{\mathscr{F}}

\newcommand{\Qsplitting}{\omega_a}

\newcommand{\MicroField}{\boldsymbol{B}}
\newcommand{\MicroAmp}{\Omega}
\newcommand{\MicroCarrierFreq}{\omega_\mu}
\newcommand{\phiNoise}{\phi_N}
\newcommand{\phiControl}{\phi_C}
\newcommand{\MicroAmpNoise}{\MicroAmp_N}
\newcommand{\MicroAmpControl}{\MicroAmp_C}

\newcommand{\sigx}{\hat{\sigma}_x}
\newcommand{\sigy}{\hat{\sigma}_y}
\newcommand{\sigz}{\hat{\sigma}_z}
\newcommand{\sigp}{\hat{\sigma}_+}
\newcommand{\sigm}{\hat{\sigma}_-}

\newcommand{\HphiN}{H_{\dot{\phi}_N}}

\usepackage{appendix}

\begin{document}
\title{Experimental bath engineering for quantitative studies of quantum control}%
\author{A. Soare}%
\author{H. Ball}
\author{D. Hayes}%
\affiliation{ARC Centre for Engineered Quantum Systems, School of Physics, The
University of Sydney, NSW 2006 Australia\\ \emph{and} National Measurement Institute, West Lindfield, NSW 2070 Australia}

\author{X. Zhen}
\affiliation{ARC Centre for Engineered Quantum Systems, School of Physics, The
University of Sydney, NSW 2006 Australia\\ \emph{and} National Measurement Institute, West Lindfield, NSW 2070 Australia}
\affiliation{\emph{also} Tsinghua University, Beijing, People's Republic of China}
\author{M. C. Jarratt}
\author{J. Sastrawan}
\affiliation{ARC Centre for Engineered Quantum Systems, School of Physics, The
University of Sydney, NSW 2006 Australia\\ \emph{and} National Measurement Institute, West Lindfield, NSW 2070 Australia}
\author{H. Uys}
\affiliation{Council for Scientific and Industrial Research, National Laser Centre, Pretoria, South Africa}
\author{M.J. Biercuk}%
\email{michael.biercuk@sydney.edu.au}
\affiliation{ARC Centre for Engineered Quantum Systems, School of Physics, The
University of Sydney, NSW 2006 Australia\\ \emph{and} National Measurement Institute, West Lindfield, NSW 2070 Australia}
\date{\today}%
%\begin{center}

%__________________________________________________________________________________________

%							     ABSTRACT 
%__________________________________________________________________________________________
 
\begin{abstract}

We develop and demonstrate a technique to engineer universal unitary baths in quantum systems.  Using the correspondence between unitary decoherence due to ambient environmental noise and errors in a control system for quantum bits, we show how a wide variety of relevant classical error models may be realized through In-Phase/Quadrature modulation on a vector signal generator producing a resonant carrier signal.  We demonstrate our approach through high-bandwidth modulation of the 12.6 GHz carrier appropriate for trapped $^{171}$Yb$^{+}$ ions.  Experiments demonstrate the reduction of coherent lifetime in the system in the presence of both engineered dephasing noise during free evolution and engineered amplitude noise during driven operations.  In both cases the observed reduction of coherent lifetimes matches well with quantitative models described herein.  These techniques form the basis of a toolkit for quantitative tests of quantum control protocols, helping experimentalists characterize the performance of their quantum coherent systems.

\end{abstract}

\maketitle

%__________________________________________________________________________________________

%							  SECTION I: INTRODUCTION 
%__________________________________________________________________________________________

\section{Introduction}\label{Sec:Intro}

The discipline of quantum control engineering~\cite{Tarn80, Tarn2003, James2007, James2009} is addressing pressing challenges in the fields of quantum physics, quantum information, and quantum engineering, attempting to provide the community with a broad range of novel capabilities in the precise manipulation of quantum systems~\cite{Foletti_NP2009, Britton2012, Lukin2013, JessenPRL2013, Islam03052013}.  For instance, protocols derived from open-loop control employing sequences of $SU(2)$ operations, known collectively as dynamical decoupling, have proven useful in extending the coherent lifetime of qubits in quantum memories~\cite{BiercukNature2009, LongStorage, RBDD, DavidsonDD} and in producing effective noise filters for quantum sensors~\cite{Bylander2011, Suter_PRL2011, Hollenberg_NP_2011, Suter_Sensing, Lukin_Nature2013, RBMagnetometry}.

Beginning with the work of Kurizki \emph{et al.}~\cite{KurizkiPRL2001, KurizkiPRL2004}, there has been a substantial effort in the field towards incorporating filter-transfer functions into the vernacular of quantum control~\cite{BiercukJPB2011, Suter_Filter}.  This has extended from trivial application of the identity in dynamical decoupling ~\cite{Martinis2003, Kuopanportti2008, BiercukNature2009,UysPRL2009,UhrigPRL2007,CywinskiPRB2008} to arbitrary single ~\cite{GreenPRL2012,UhrigPRA2012, GreenNJP2013, Kabytayev2014} and two-qubit operations ~\cite{HayesPRL, HayesPRA2011}.  In this framework, a metric of interest - generally an ensemble-averaged operational fidelity - may be simply calculated from the product of the environmental noise power spectrum and a filter transfer function capturing the effects of the control in the Fourier domain.  This approach has been shown to be a general and efficient approach capturing arbitrary control and arbitrary universal noise in quantum systems~\cite{GreenNJP2013} and is a powerful tool for understanding the influence of realistic coloured classical noise power spectra on quantum systems.

These advances are providing a means for theoretical researchers to move away from the unphysical Markovian assumptions for stochastic, uncorrelated error models selected for convenience in quantum error correction and the like ~\cite{NC, QuirozPRA2009}, and has provided a simple platform for the development of novel protocols aimed at improving control fidelity in quantum systems.  As these protocols transition from theoretical concepts into the laboratory, experimentalists require techniques to quantitatively verify the predicted performance in different noise environments and compare outcomes in a manner that is insensitive to underlying imperfections in their hardware.  Such precise validations are necessary for researchers to confidently develop quantum control techniques using substantiated methodologies and subroutines.

In this manuscript we describe a technique to engineer arbitrary unitary baths consisting of dephasing and amplitude damping processes for quantitative tests of experimental quantum control.  We present a simple theoretical model for approximating arbitrary classical power spectra via discrete frequency combs with user-selected envelopes (e.g. $1/f$).  We describe how this model permits simple and verifiable creation of time-dependent noise realizations in both dephasing and amplitude-damping quadratures, compatible with experimental systems.  Through demonstration of the isomorphism of unitary control errors and environmental decoherence we map these noise realisations to modulation of a carrier signal in an experimental control system, \emph{e.g.} for a single quantum bit.  Using trapped $^{171}$Yb$^{+}$ ions with splitting $\sim$12.6 GHz, we demonstrate our bath engineering approaches via $IQ$ modulation on the microwave carrier.  Ramsey spectroscopy measurements quantitatively verify the predicted influence of engineered dephasing noise on the coherent lifetime of our qubits.

The remainder of this paper is structured as follows. In Section~\ref{Sec:Theory}, we provide a detailed theoretical derivation of our selected method of unitary noise engineering for both dephasing (detuning) and amplitude damping Hamiltonians, and describe how these noise spectra may be translated to widely available time-domain $IQ$-modulation waveforms applied to a carrier signal.  We then move on to describe our experimental system and its capabilities in in Section~\ref{Sec:Our Model Platform}.  This is followed in Section~\ref{Sec:Experiment} by a characterization of an experimentally implemented noise-engineered bath through direct examination of the carrier and a demonstration of engineered dephasing environments via measurements of coherent lifetimes for $^{171}$Yb$^{+}$ ion qubits.  The manuscript concludes with a discussion and outlook towards future experiments.

%__________________________________________________________________________________________

%		      SECTION II: Physical Setting
%__________________________________________________________________________________________

\section{Physical Setting}\label{Sec:Theory}
In many quantum systems of interest we may consider two general classes of unitary \emph{time-dependent} errors. Dephasing processes are associated with rotations about $\hat{\sigma}_z$ induced by a stochastic \emph{relative} detuning between a qubit's transition (angular) frequency $\Qsplitting $ and the experiment's master clock, defined by a local oscillator (LO).  Dephasing is frequently dominated by instabilities in the qubit splitting  caused by environmental (e.g. magnetic field) fluctuations.  However, in the limit of very stable qubits (e.g. clock transitions in atomic systems~\cite{Langer2005, OlmschenkPRA2007}), observed dephasing may be caused by frequency instabilities in the experimental LO.   Similarly, one may consider coherent amplitude damping processes, causing unwanted rotations along meridians of the Bloch sphere, and arising either through ambient environmental fluctuations (e.g. microwave leakage from nearby systems) or from imperfections in the amplitude of the applied control field.  

Together these two classes of error capture so-called \emph{universal} (multi-axis) rotations of the Bloch sphere.  Importantly, the consideration of time-dependent errors in both dephasing and amplitude quadratures allows us to capture the dominant forms of non-Markovian noise processes characterized by the presence of long-time correlations; realistic laboratory settings are typically dominated by such noise terms.  Dissipative error pathways with Markovian characteristics may be captured through linearly independent error terms that we ignore through this treatment, as quantum control generally provides no relevant benefits in error resistance for these effects.

We consider a model quantum system consisting of an ensemble of identically prepared noninteracting qubits immersed in a weakly interacting noise bath and driven by an external control device. Working in the interaction picture with respect to the qubit splitting, state transformations are represented as unitary rotations of the Bloch vector. Including both control and noisy interactions, we may therefore write the generalized time-dependent Hamiltonian 
\begin{equation}\label{Eq:ham0}
\Htot(t)=\Hc(t)+ \Herr(t).
\end{equation}
The term $\Hc(t)=\mathbf{h}(t)\boldsymbol{\sigma}$ represents perfect control over the qubit state via the application of an external field, while the generalized noise term $\Herr(t)=\boldsymbol{\eta}(t)\boldsymbol{\sigma}$ captures all interactions due to the noise bath. Here $\boldsymbol{\sigma}$ denotes a column vector of Pauli matrices and the row vectors $\mathbf{h}(t),\boldsymbol{\eta}(t)\in\mathbb{R}^3$ denote respectively the Cartesian components of the control and noise fields in the basis of Pauli operators ~\cite{UhrigJPA2008,UhrigPRA2012, GreenNJP2013}. The stochastic noise fields $\eta_i(t)$, $i\in\{x,y,z\}$ model semi-classical time-dependent error processes in each of the three spatial directions.  In this formulation, dephasing processes are captured through the appearance of stochastic terms along $\hat{\sigma}_{z}$. General coherent amplitude damping terms on the other hand are captured by terms proportional to the spin operator $\hat{\sigma}_\phi:=\cos(\phi)\hat{\sigma}_x+\sin(\phi)\hat{\sigma}_y$ parametrized by the driving phase $\phi\in[0,2\pi]$.  

Our choice to write separate control and noise terms in the Hamiltonian in Eq. \ref{Eq:ham0} belies the fact that, when expressed in an appropriate interaction picture, time-dependent fluctuations in either term are effectively indistinguishable.  This observation permits a formulation in which the noise terms are all incorporated into the control Hamiltonian, and one assumes the presence of a perfectly stable qubit (\emph{i.e.} there is no ambient decoherence). This is a good approximation in the case of a sufficiently stable qubit so long as native error rates and ambient noise susceptibilities are small compared to relevant scales under study.  We therefore proceed by providing a model for quantum dynamics that permits us to capture unitary decoherence through the control.  %That is, we simply absorb the generalized noise fields into the corresponding components of the \emph{ideal} control. 

With these generalized notions in mind we proceed in laying out the detailed Hamiltonian framework relevant to our study.  The system considered in this paper consists of a qubit with transition (angular) frequency $\Qsplitting$, driven by a LO with magnetic field component aligned with $\sigz$ taking the form 
\begin{align}
\label{MicrowaveField}\MicroField(t) &= \MicroAmp(t)\cos(\MicroCarrierFreq t+\phi(t))\hat{\mathbf{z}}\\
\label{MicrowaveAmplitude}\MicroAmp(t) &= \MicroAmpControl(t)+\MicroAmpNoise(t)\\
\label{MicrowavePhase}\phi(t) &= \phiControl(t)+\phiNoise(t)
\end{align}
with $\MicroCarrierFreq$ the carrier frequency.  In this formulation the time-dependence of the phase $\phi(t)$ and amplitude $\MicroAmp(t)$ has been formally partitioned into components denoted by the subscripts $C$ and $N$, capturing the desired control and noisy interactions respectively.  Using standard approximations, and working in an interaction picture (see Appendix~\ref{Sec:AppendixA}), the system Hamiltonian ($\hbar=1$) may be expressed
\begin{align}\label{PhiNnteractionHamiltonian}
H_I= -\frac{\dot{\phi}_N(t)}{2}\sigz +
\frac{1}{2}\Omega(t)
\Big\{
\cos[\phi_C(t)]\sigx+
\sin[\phi_C(t)]\sigy
\Big\}.
\end{align}
\noindent The noise component $\phi_N(t)$ of the engineered phase $\phi(t)$ produces net rotations about $\sigz$ through its time derivative, $\dot{\phi}_N(t)$, and the resonant carrier field drives coherent Rabi flopping between the qubit states $\ket{1}$ and $\ket{0}$. The instantaneous Rabi rate in this case is proportional to $\MicroAmp(t)$ and rotations, generated by the spin operator $\hat{\sigma}_{\phi_{C}(t)}$, are driven about the axis $\vec{r}=\left(\cos\phi_C(t), \sin\phi_C(t),0\right)$ in the $xy$-plane of the Bloch sphere.

Given sufficient control over both the phase and amplitude of our driving field, Eq. \ref{PhiNnteractionHamiltonian} therefore indicates we may engineer a variety of effective control Hamiltonians with dephasing and amplitude damping terms of interest. For instance, setting $\phi_C(t) = 0$, we may generate 
%\begin{equation}
%\hat{H}_0\propto
%\begin{cases}
%h_{x}\left(1+\eta_{x}(t)\right)\hat{\sigma}_{x}\;&\text{Multiplicative amplitude noise}\\
%\left(h_{x}+\eta_{x}(t)\right)\hat{\sigma}_{x}\;&\text{Additive amplitude noise}\\
%h_{z}\left(1+\eta_{z}(t)\right)\hat{\sigma}_{z}\;&\text{Additive dephasing noise}
%\end{cases}
%\end{equation}
\begin{equation}\label{Eq: EngineeredHamiltonians}
\hspace{-0.19cm}\Htot(t)\propto
\begin{cases}
h_{x}(t)\left(1+\eta_{x}(t)\right)\hat{\sigma}_{x}\;&\text{(\emph{Mult. Amp. Noise})}\\
\left(h_{x}(t)+\eta_{x}(t)\right)\hat{\sigma}_{x}\;&\text{(\emph{Add. Amp. Noise})}\\
h_{x}(t)\hat{\sigma}_x+\eta_z(t)\hat{\sigma}_z\;&\text{(\emph{Add. Deph. Noise})}
\end{cases}
\end{equation}
where the control field $h_{x}(t)$ is proportional to $\Omega_C(t)$ and the noise fields $\eta_x(t)$ or $\eta_z(t)$ may be switched on, with desired spectral properties, by an appropriate choice of $\Omega_N(t)$ and $\dot{\phi}_N(t)$ respectively. 

The Hamiltonians in Eq. \ref{Eq: EngineeredHamiltonians} correspond to familiar error models from NMR and quantum information~\cite{Vandersypen2004,MerrillArXv2012}, but now explicitly incorporate non-Markovian time-dependent effects through the power spectra of the relevant terms in $\boldsymbol{\eta}(t)$.  The first noise model may be produced in the absence of Hamiltonian terms that look like $h_{x}(t)\eta_{x}(t)$ by virtue of the ability to arbitrarily parameterize $\eta_{x}\{h_{x}(t),t\}$.  

Following previous work we express the first-order averaged fidelity of an arbitrary unitary control operation on $SU(2)$ in the presence of noise as~\cite{GreenNJP2013} 
\begin{align}
\mathcal{F}_{av}(\tau)&\approx\frac{1}{2}\big\{1+\exp[-\chi(\tau)]\big\}\\
%\chi(t)&=\frac{2}{\pi}\Big[\int_{0}^{\infty}\frac{d\omega}{\omega^{2}}S_{x}(\omega)F_{x}(\omega)+\\
%&\int_{0}^{\infty}\frac{d\omega}{\omega^{2}}S_{y}(\omega)F_{y}(\omega)+\\
%&\int_{0}^{\infty}\frac{d\omega}{\omega^{2}}S_{z}(\omega)F_{z}(\omega)\Big]\\
\label{Eq: DecoherenceChi}\chi(\tau)&=\frac{2}{\pi}\Big[
\int_{0}^{\infty}
\frac{d\omega}{\omega^2}
\Sd(\omega)
\Fd(\omega)+\nonumber\\
&\hspace{0.825cm}\int_{0}^{\infty}
\frac{d\omega'}{\omega'^2}
\Sa(\omega') 
\Fa(\omega')\Big].
\end{align}
%\begin{align}
%\mathcal{F}_{av}(t)&\approx\frac{1}{2}\left[1+\exp{-\chi(t)}\right]\\
%\chi(t)&=\frac{2}{\pi}\Big[\int_{0}^{\infty}\frac{d\omega}{\omega^{2}}S_{x}(\omega)F_{x}(\omega)+\\
%&\int_{0}^{\infty}\frac{d\omega}{\omega^{2}}S_{y}(\omega)F_{y}(\omega)+\\
%&\int_{0}^{\infty}\frac{d\omega}{\omega^{2}}S_{z}(\omega)F_{z}(\omega)\Big].
%\end{align}

\noindent Here we have defined independent noise power spectra $\Sd(\omega)$ and $\Sa(\omega)$, with angular frequency $\omega$, for fluctuations in $\dot{\phi}_N(t)$ and $\Omega_N(t)$ respectively, while the quantities $\Fd(\omega)$ and $\Fa(\omega)$ represent the spectral characteristics of the control under study.  While we will not focus on the particular form of this so-called filter transfer function expression for operational fidelity~\cite{GreenPRL2012, GreenNJP2013}, we can clearly see the importance of these noise power spectra in determining the performance of an arbitrary control operation. Consequently, in the following section, we derive their forms.
\bla
%__________________________________________________________________________________________

%		      SECTION III: Engineering Noise in the Control System
%__________________________________________________________________________________________

\section{Engineering noise in the control system}

In the laboratory we rely on engineering noise in our control system to provide a method to accurately reproduce decoherence processes of interest.  This approach has significant benefits over e.g. noise injection in ambient magnetic field coils, as it minimises potential nonlinearities and frequency-dependent responses in hardware elements, exploiting instead the modulation capabilities of a carrier synthesis system~\cite{Pozar}.  By engineering noise through a highly accurate control system with linear response we gain the ability to perform quantitative tests of quantum control in the presence of unitary noise Hamiltonians.  %In particular, we are able to engineer a variety of well-known error models incorporating time-dependent noise processes. 

We employ the  phase- and amplitude-modulation capabilities in state-of-the-art quantum control systems in order to provide access to the error models of interest.  In the remainder of this section we present a mathematical formalism linking our error model in the geometric picture of unitary dynamics to the properties of a near-resonant drive field of the form given in Eq. \ref{MicrowaveField}.
%%%%%\red 
%%%%%\begin{align}
%%%%%\label{MicrowaveField}\MicroField(t) &= \MicroAmp(t)\cos(\MicroCarrierFreq t+\phi(t))\\
%%%%%\label{MicrowaveAmplitude}\MicroAmp(t) &= \MicroAmpControl(t)+\MicroAmpNoise(t)\\
%%%%%\label{MicrowavePhase}\phi(t) &= \phiControl(t)+\phiNoise(t).
%%%%%\end{align}
%%%%%Here $\MicroCarrierFreq$ denotes the carrier frequency, and the time-dependence of the phase $\phi(t)$ and amplitude $\MicroAmp(t)$ under our modulation scheme has been made explicit. Both quadratures are formally partitioned into the desired control and added noise components, denoted by the subscripts $C$ and $N$ respectively. 
%%%%%\bla
%%%%%
%%%%%\red Setting $\MicroCarrierFreq = \Qsplitting$ the resonant carrier field in Eq. \ref{MicrowaveField} drives coherent rotations between the qubit basis states. In this case, the Rabi rate is proportional to $\MicroAmp(t)$ and rotations, generated by the spin operator $\hat{\sigma}_{\phi(t)}$, are driven about the axis $\vec{r}=\left(\cos\phi(t), \sin\phi(t),0\right)$ in the $xy$-plane of the Bloch sphere.  For instance, a noise-free $\pi$ pulse about the $x$-axis would have $\Omega_{N}=\phi_{N}=0$, and $\phi_{C}=0$, $\Omega_{C}(t)=\Omega$ for $t\in[0,\tau_{\pi}]$, with $\tau_{\pi} = \pi/\Omega$.
%%%%%\bla

%				---------------------------------
				%subSECTION: ENGINEERED DETUNING NOISE 
%				---------------------------------

\subsection{Arbitrary dephasing (detuning) power spectra}
We begin with the case of noise proportional to $\hat{\sigma}_{z}$. Our method relies on generating stochastic detuning errors by performing \emph{phase modulation} on a constant-amplitude carrier, thereby implementing an effective pure dephasing Hamiltonian. Setting $\MicroAmpNoise(t) = \phiControl = 0$ and $\MicroAmpControl = \Omega_0$ we write
\begin{align}\label{DephasingMicrowave}
\MicroField(t) &= \Omega_0\cos(\omega_\mu t+\phiNoise(t))\hat{\mathbf{z}}\\
\phiNoise(t) &= \alpha\sum_{j=1}^JF(j)\sin(\omega_j t+\psi_j)
%\omega_j &= j\omega_0
\end{align}
where $\psi_j$ is a random number. That is, the driving carrier tone is modulated to include a time-dependent, stochastic error in the \emph{phase} constructed as a discrete Fourier series with a base frequency $\omega_0=\omega_j/j$, with $\alpha$ being a global scaling factor~\cite{NoiseGen}.

%with spectral components that look like a discretized Fourier synthesis of sinusoids.  This discrete Fourier decomposition is based on harmonics of a base frequency expressed as $\omega_j=j\omega_{0}$ where each tooth in the comb is spectrally weighted by $\alpha F(j)$, with $\alpha$ is a global scaling factor. 

The link between this phase modulation and the dephasing power-spectral density of interest is revealed by defining the instantaneous phase in terms of the carrier plus a time-dependent detuning $\Bd(t)$
\begin{align}\label{InstantaneousPhase}
\Phi(t)=\Phi_0+\int_0^tdt'\Big[\omega_\mu+\Bd(t')\Big]
\end{align}
\noindent where $\Bd(t)$ is a zero-mean time-dependent random variable.  This then implies 
\begin{align}
\phiNoise(t) = \Phi_0+\int_0^td\tau\Bd(\tau)\hspace{0.25cm}\iff\hspace{0.25cm}\Bd(t) = \frac{d}{dt}\phiNoise(t)
\end{align}
so the time-dependent detuning noise $\Bd(t)$, explicitly linked to the phase modulation of the carrier, characterizes the strength of the dephasing noise term in Eq. \ref{PhiNnteractionHamiltonian}.

%The link between this phase modulation and the dephasing power-spectral density of interest is revealed by defining the instantaneous frequency $\omega(t):=\MicroCarrierFreq+\Bd(t)$ in terms of the carrier plus a detuning specified by the zero-mean time-dependent random variable $\Bd(t)$. In this case the instantaneous phase $\Theta(t)=\Theta_0+\int_0^tdt'\omega(t')$ is given by 
%\begin{align}\label{InstantaneousPhase}
%\Theta(t)=\Theta_0+\int_0^tdt'\Big[\omega_\mu+\Bd(t')\Big].
%\end{align}
%But from Eq. \ref{DephasingMicrowave} we know $\Theta(t)=\omega_\mu t+\phiNoise(t)$. Substituting into Eq. \ref{InstantaneousPhase} we therefore obtain 
%\begin{align}
%\phiNoise(t) = \Theta_0+\int_0^td\tau\Bd(\tau)\hspace{0.25cm}\iff\hspace{0.25cm}\Bd(t) = \frac{d}{dt}\phi(t).
%\end{align}

 Using the Euler decomposition we may then write
%\begin{align}
%\beta(t) =\alpha\omega_0\sum_{j=1}^JjF(j)\cos(\omega_j t+\psi_j),
%&=\alpha\sum_{j=1}^JF(j)\omega_j\cos(\omega_j t+\psi_j)\\
%\end{align}
%or expanding in Euler terms
\begin{align}\label{DetuningNoise}
\Bd(t) =\frac{\alpha\omega_0}{2}\sum_{j=1}^JjF(j)\Big[e^{i(\omega_j t+\psi_j)}+e^{-i(\omega_j t+\psi_j)}\Big].
\end{align}
\noindent Assuming wide-sense stationarity, the two-time correlation function for $\Bd(t)$ is then written
\begin{equation}
\langle\Bd(t+\tau)\Bd(t)\rangle_t=\frac{\alpha^2\omega_0^2}{2}\sum_{j=1}^J(jF(j))^2\cos(\omega_j \tau)
\end{equation}
\noindent  where $\langle\cdot\rangle_t$ denotes averaging over all times $t$ from which the relative lag of duration $\tau$ is defined. Invoking the Wiener-Khintchine theorem~\cite{MillerBook2012} and moving to the Fourier domain we then obtain the power spectral density
\begin{align}\label{PSDCombMain}
\Sd(\omega) = \frac{\pi\alpha^2\omega_0^2}{2}\sum_{j=1}^J(jF(j))^2\Big[\delta(\omega-\omega_j)+\delta(\omega+\omega_j)\Big].
\end{align}
\noindent The detailed derivation of the above expressions appears in Appendix \ref{ApdxSec: PSDs}.

The power-spectral density is thus represented as a Dirac comb of discrete frequency components with the amplitude of the $j$th tooth determined by the quantity $(j(F(j))^2$.  We now have an explicit relationship between an effective \emph{dephasing} or \emph{frequency detuning} noise power spectrum and the phase modulation of the carrier frequency in the control system required to achieve that PSD.

It is then straightforeward to specify the construction of any power-law PSD by writing the amplitude of the $j$th frequency component as a power-law, $\Sd (\omega_j)\propto (j\omega_{0})^{p}$. It therefore follows that the envelope function for the comb teeth in the phase modulation scales as
\begin{eqnarray}\label{FunctionalFormOfFjForPowerLawpNoise}
F(j) = j^{\frac{p}{2}-1}.
\end{eqnarray}
Table~\ref{NoisePowerLaws} shows the functional form required for $F(j)$ in order to achieve dephasing-noise PSDs of interest. 

%\begin{eqnaray}
%F(j) = \frac{\sqrt{P(j)}}{j}
%\end{eqnarray}
%In particular, suppose $P(j)$ takes the simplest
%\begin{eqnarray}
%P(j) = j^p,
%\end{eqnarray} 

%				---------------------------------
				%subSECTION: ENGINEERED AMPLITUDE  NOISE 
%				---------------------------------

\subsection{Arbitrary amplitude power spectra}
Derivation of the relevant amplitude noise power spectra proceeds in a similar manner.  We consider here multiplicative amplitude noise, although the derivation maintains a similar form in the case of additive noise.  Further, in our model the amplitude noise is always assumed to be coaxial with the driving field.  While this is not a strict requirement, it greatly simplifies the analysis and broadly represents an interesting class of time-dependent error models incorporating driving-field noise. The relevant modulation capability here is, as expected, amplitude modulation on a carrier signal. Setting $\phi(t) = 0$, $\MicroAmpControl(t) = \Omega_0$ and $\MicroAmpNoise = \Omega_0\Ba(t)$ Eq. \ref{MicrowaveField} reduces to 
\begin{align}\label{GeneralizedPhaseNoise}
\MicroField(t) &= \big(\Omega_0(1+\Ba(t))\big)\cos(\omega_{\mu} t)\hat{\mathbf{z}}.
\end{align}
That is, amplitude modulation transforms the control field strength as
\begin{align}\label{AmpNoise}
\Omega_0\rightarrow\Omega_0(1+\Ba(t))
\end{align}
where again, $\Ba(t)$ is a zero mean stochastic random variable here capturing fluctuations in the drive amplitude.  This term is realized directly through the comb of discrete frequency components with randomly selected phase shifts
\begin{align}\label{GeneralizedAmplitudeNoise}
\Ba(t) &= \alpha\sum_{j=1}^JF(j)\cos(\omega_j t+\psi_j)\\
&=\frac{\alpha}{2}\sum_{j=1}^JF(j)\Big[e^{i(\omega_j t+\psi_j)}+e^{-i(\omega_j t+\psi_j)}\Big].
\end{align}
\noindent The form of this expression is similar to that above for dephasing noise, except the direct amplitude modulation removes a factor of $j$ from the expression.  We are interested in producing a PSD for the quantity $\Ba(t)$ as this captures the amplitude errors pertaining to the second term of Eq. \ref{Eq: DecoherenceChi}. Following the same method as before (see Appendix \ref{ApdxSec: PSDs}) we obtain 

\begin{align}\label{AmplitudePSDComb}
\Sa(\omega) &= \frac{\pi\alpha^2}{2}\sum_{j=1}^J(F(j))^2\Big[\delta(\omega-\omega_j)+\delta(\omega+\omega_j)\Big].
\end{align}
Once again we may define a relationship between the power-law of the target noise power spectral density and the quantity $((F(j))^2$, which determines the amplitude of the $j$th tooth in the frequency comb PSD above.   In this instance the removal of a differential relationship between the modulation and desired noise power spectrum yields the simplified expression $F(j)=j^{p/2}$ for $\Sa(\omega_j)\propto(j\omega_{0})^{p}$.

\subsection{Summary}

With these relationships we now have explicit links between the quantities we wish to engineer in realizing unitary dephasing or relaxation noise power spectra and the relevant parameters entering into the modulation of a control signal. We will employ these relations to engineer arbitrary unitary noise baths for quantitative tests of various quantum control protocols.

\begin{figure}[tp]
\centering
\includegraphics[width=7cm]{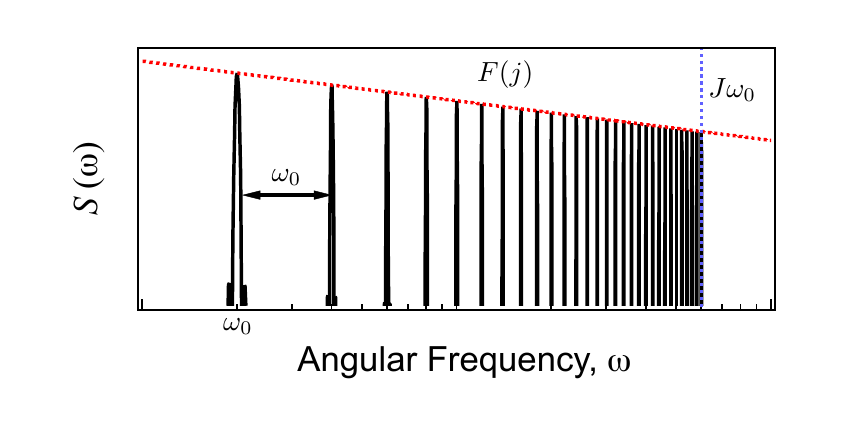}\\
\caption{(color online) Schematic depiction of the frequency comb generated in our noise engineering protocol and its relationships to key parameters on a logarithmic scale.
\label{Fig:Comb}}
\end{figure}

In implementing bath engineering in the laboratory we are left with the following free-parameters:
\begin{itemize}
\item $z/\Omega$: The quadrature of noise injection (dephasing vs amplitude)
\item $\omega_{0}$: The fundamental frequency of the Dirac comb and the \emph{lower-cutoff} of the noise power spectrum
\item $J$: The maximum number of comb teeth in the discrete sum, setting the upper frequency cutoff $J\omega_{0}$
\item $p$: The exponent setting the frequency dependence of the effective noise power spectrum
\end{itemize}
\noindent These parameters provide an experimentalist with a broad set of capabilities for bath engineering (see Table~\ref{NoisePowerLaws} and Fig.~\ref{Fig:Comb}).

\begin{table}[htp]\label{}
  \centering
\begin{tabular}{c|cccc||cccc|}
\cline{2-9}
 & \multicolumn{4}{|c||}{Dephasing}  &\multicolumn{4}{c|}{Amplitude}\\
\cline{2-9}
 &$1/f^2$ &  $1/f$ &White &Ohmic &$1/f^2$ &  $1/f$ &White &Ohmic  \\
\hline
\multicolumn{1}{|c|}{p}& $-2$ &$-1$ & $0$  &  $1$ & $-2$ &$-1$ & $0$  &  $1$  \\
\multicolumn{1}{|c|}{$F(j)$}& $j^{-2}$ &$j^{-3/2}$ & $j^{-1}$  &  $j^{-1/2}$& $j^{-1}$ &$j^{-1/2}$ & $j^{0}$  &  $j^{1/2}$\\
\hline
\end{tabular}
\caption{Functional form of $F(j)$ for well-known dephasing and amplitude noise PSDs.}\label{NoisePowerLaws}
\end{table}

%__________________________________________________________________________________________

%		      		SECTION IV: EXPERIMENTAL BATH ENGINEERING
%__________________________________________________________________________________________

\section{Experimental Bath Engineering}\label{Sec:Experiment}

%				---------------------------------
				%subSECTION: Experimental Platform
%				---------------------------------
\subsection{Experimental Platform}\label{Sec:Our Model Platform}

\begin{figure*}[htp]
\centering
\includegraphics[width=17cm]{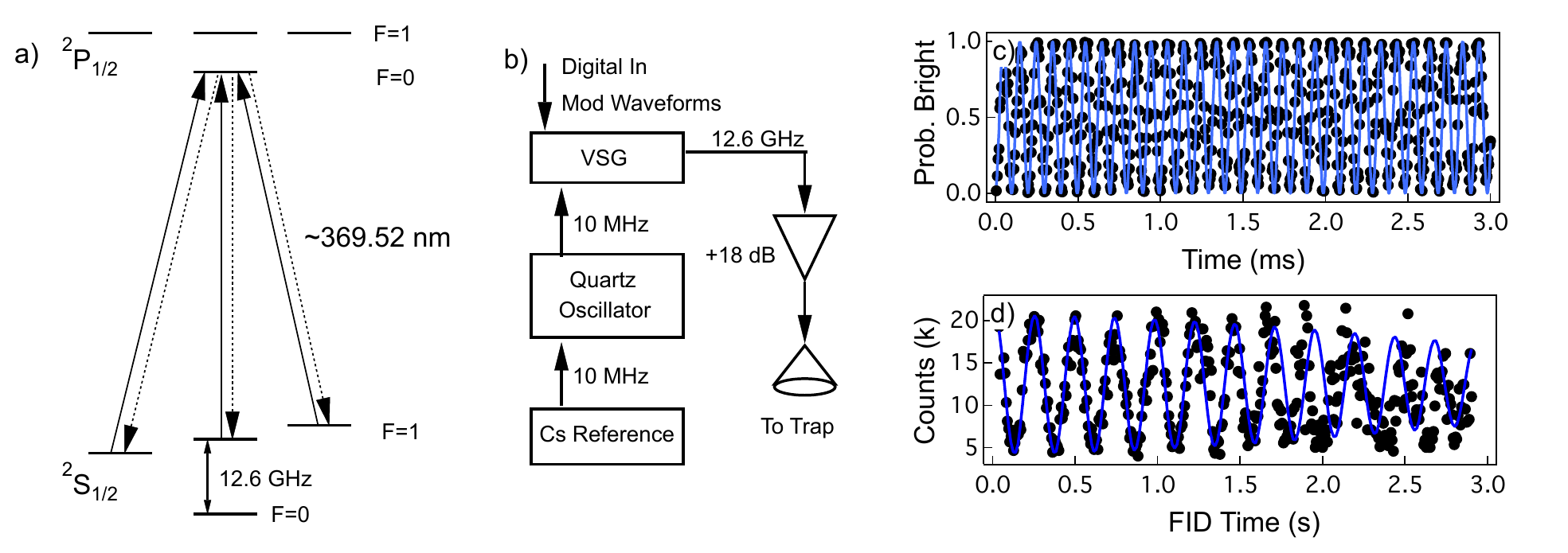}\\
\caption{(color online) Experimental control system.  a) Simplified level structure of $^{171}$Yb$^{+}$ ions with qubit at 12.6 GHz splitting highlighted.  Solid arrows indicate excitation via UV detection laser, dotted arrows indicate spontaneous emission pathways.  Repumping transitions to mestatable D and F states not shown.   b) Microwave synthesis chain employed for $^{171}$Yb$^{+}$ qubits.  Digital programming of the vector signal generator (VSG) conducted via either GPIB or LAN.  c) Measured Rabi flopping at 12.6 GHz on the clock transition in the $^{171}$Yb$^{+}$ ground state.  In this measurement we use Bayesian estimation to map raw measured photon counts to bright or dark state probability.  d) Free evolution measured via Ramsey interferometry and presented using raw photon counts for simplicity.  Interrogating $\pi/2$ pulses applied with frequency detuning of $\sim4$ Hz to yield interference fringes.  After approximately two seconds of free evolution the fringe frequency appears to shift abruptly and then become unstable.  The overlaid damped oscillation assumes a Gaussian decay and a $T_{2}$ of four seconds and matches the general decay envelope well.  However, due to the appearance of statistically non-stationary dynamics during the experiment, fitting struggles to provide an accurate reproduction of the data.  Nonetheless, the data clearly indicate coherence between the LO and qubits beyond approximately three seconds. 
\label{Fig:ExptSchem}}
\end{figure*}

The approach we have described above is quite generic for the case of a quantum system controlled by an oscillatory signal, including most atomic, superconducting, and many semiconductor-based spin qubits.  In this section we describe the experimental platform we will employ for validation of our method.

We use trapped \textsuperscript{171}Yb\textsuperscript{+} ions as our model experimental platform; a detailed description of related experimental approaches appears in~\cite{OlmschenkPRA2007}. Neutral \textsuperscript{171}Yb is ionized using a two photon process whereby 399 nm light excites electrons from \textsuperscript{1}S$_{0}$ to \textsuperscript{1}P$_{1}$ and 369 nm light is sufficiently energetic to further excite electrons to the continuum. A linear Paul trap enclosed in an ultra-high vacuum (UHV) chamber is used to trap several hundred \textsuperscript{171}Yb\textsuperscript{+} ions.  Doppler cooling of the ions is achieved using 369 nm laser light, slightly red-detuned from the \textsuperscript{2}S$_{1/2}$ to \textsuperscript{2}P$_{1/2}$ transition. Additional lasers near 935 nm and 638 nm are employed to depopulate metastable states.  Typical experiments employ ensembles of approximately 100-1000 ions with high-homogeneity in magnetic field and microwave field over the ensemble.

Our qubit is the 12.6 GHz hyperfine splitting between the \textsuperscript{2}S$_{1/2}\ket{F=0,m_{f}=0}$ and \textsuperscript{2}S$_{1/2}\ket{F=1,m_{f}=0}$ states. For notational simplicity we denote these states by $\ket{0}$ and $\ket{1}$ respectively. The system may be optically pumped to $\ket{0}$ using a 2.1 GHz sideband on the 369 nm cooling beam, which couples the states \textsuperscript{2}S$_{1/2}\ket{F=1}\leftrightarrow$\textsuperscript{2}P$_{1/2}\ket{F=1}$ following ~\cite{OlmschenkPRA2007}.  

State detection is achieved using resonant light near 369nm, which preferentially couples the state $\ket{1}$ to the excited $P$-state, resulting in a large probability of detecting scattered photons.  These photons are detected using a pair of large-diameter lenses and a photomultiplier tube.  State discrimination is conducted by photon counting followed by conversion to a probability that the Bloch vector lies at a particular location along a meridian of the Bloch sphere. This measurement is susceptible ion loss in the ensemble and both laser amplitude and frequency drifts over long timescales, resulting in variable maximum and dark count rates over time.  We therefore employ a normalization and Bayesian estimation procedure to improve measurement fidelity.

Dark state normalization is achieved by cooling and optically pumping the ions to the $F=0$ state of the 12.6 GHz hyperfine manifold and then performing a measurement of photon counts, while bright state normalization includes an additional $\pi_{x}$ gate implemented using microwaves before the photon count measurement. We have experimentally ascertained that for any angle of declination of the Bloch vector with respect to the $-z$ axis, $\theta$, photon counts over a repeated number of identical experiments are normally distributed about a mean value with standard deviation dominated by rapid laser frequency fluctuations with magnitude $\sim1$ MHz.

We use Bayesian inference to statistically determine the qubitÕs $z$-projection given the number of scattered 369 nm photons we measure, denoted $P(\theta|c)$. We write
\begin{align}
P(\theta|c)&=\frac{P(c|\theta)P(\theta)}{P(c)}\\
&=\frac{P(c|\theta)P(\theta)}{ \sum_i\int_{0}^{\pi}d\theta P(c|\theta)P(\theta)}
\end{align}
where in the second line we have incorporated the fact that $P(c)$ is dependent on our knowledge of the probability distribution function for $\theta$. We have experimentally verified that we may write the probability density function $P(c|\theta)=\exp(-(\frac{D+\frac{B-D}{\pi}\theta}{ \sigma_{D}+\frac{\sigma_{B}-\sigma{D}}{\pi}\theta })^{2})$. Here we have defined a Gaussian with centre defined by linearly interpolating between the mean detected photon counts for the bright state, $B$, and mean counts for the dark state, $D$. The standard deviation of the Gaussian is defined by linearly interpolating between the standard deviations of the photon count distributions for the bright state $\sigma_{B}$, and dark state $\sigma_{D}$. The quantities $B$, $D$, $\sigma_{B}$, and $\sigma_{D}$ are found by performing bright and dark state normalization before each iteration of the experiment of interest.  $P(\theta)$ is initially assumed to be a uniform probability distribution, and is iteratively refined with subsequent experiments. We then calculate the mean and standard deviation and apply the transform $P_{\ket{1}}=\sin^2(\frac{\theta}{2})$.  Using this method we can achieve measurement fidelity in excess of $98\%$.

A simpler method of calculating the approximate bright state probability is to take a simple normalized average of the form $(E-D)/(B-D)$ where $B$ and $D$ are define as before and $E$ is the mean detected photon counts for the experiment of interest.   This method agrees well with Bayesian estimation for states that are not near the poles of the Bloch sphere and provides a simpler and faster measurement method in cases where maximizing measurement fidelity near the poles is not required.

To produce our master oscillator signal we use an ultra-low phase noise vector source referenced to a Caesium clock and 10 MHz Wenzel cleanup-oscillator for long term stability and good short-term phase noise.  The output of the signal generator is amplified using a low-phase-noise amplifier with maximum output of approximately $+33$ dBm.   A commercially available, microwave horn-lens combination is used to produce a highly directional free-space linearly polarized microwave field ($+25$ dBi directional gain) which can be directed at the ions, approximately 150~mm from a 150~mm diameter viewport on the UHV chamber.  

Coherent rotations between the measurement basis states are driven by using the magnetic field component of resonant microwave radiation.  The Rabi rate for driven oscillations is linearly proportional to the microwave magnetic field amplitude, with rotations about an axis $\vec{r}$ lying on the $xy-$plane of the Bloch sphere and set by the phase of the microwaves as $\vec{r}=\left(\cos\phi(t), \sin\phi(t),0\right)$.  Rabi flopping experiments (Fig.~\ref{Fig:ExptSchem}c) demonstrate high-visibility coherent qubit rotations where we can achieve hundreds of flops before seeing appreciable decay.  With $\sim+30$ dBm nominal microwave power (\emph{e.g.} not accounting for cable losses) we achieve $\pi$-times as low as $\sim15\;\mu$s, but we typically operate near $50\;\mu$s.  We have confirmed that in these experiments our measured Rabi flopping times are limited by small microwave field amplitude inhomogeneities over the ion ensemble caused by diffraction of the microwave beam at the aperture of the UHV chamber. %Based on studies of Rabi flopping on the $\sigma_{\pm}$ transitions in $^{171}$Yb$^{+}$ we find the microwave system (as aligned) delivers linear to circular polarisation at a power ratio of approximately XXX.  

A standard technique for characterising oscillator stability is Ramsey spectroscopy~\cite{Ramsey1956}.  We prepare the ions in $\ket{0}$ and rotate to $\ket{+y}$ using a $\pi/2$-pulse applied about $\hat{x}$, but slightly detuned from resonance by +4 Hz.  After a free evolution period, a second $\pi/2$ pulse will rotate the qubit to $\ket{0}$ or $\ket{1}$ depending on the phase accumulated between the master oscillator and the qubit.  Scanning the evolution time, $\tau$, reveals sinusoidal fringes due to the free evolution of the qubit relative to the control during the delay period. Instabilities of the phase over time cause the relative phase between the qubit and master oscillator to become randomised, thus reducing the visibility of Ramsey fringes.  

An important advantage of this system is that the selected qubit transition is first order insensitive to magnetic field fluctuations. As a result the intrinsic free-evolution coherence time of this hyperfine qubit has been measured to be at least 15 minutes ~\cite{FiskIEEE1997}.    A $T_{2}$ decay time of approximately four seconds, inferred from Ramsey experiments (Fig. \ref{Fig:ExptSchem}d), demonstrates long term coherence between the qubit and our LO, ultimately limited by phase stability of the LO (typically $-80$ dBc phase noise at 100 Hz offset from carrier).  These experimental measurements reveal that this system therefore provides a ``clean'' baseline for quantitative tests of bath engineering.

%				---------------------------------
				%subSECTION: Implementation of bath engineering by IQ mod
%				---------------------------------
\subsection{Implementation of bath engineering by \emph{IQ} Modulation}
The bath engineering technique described above provides a generic framework allowing noise to be generated for specific Hamiltonians of interest.  We must now demonstrate how such noise may be implemented using the kind of control hardware typically available for quantum control experiments: $IQ$ modulation on the resonant carrier.

\begin{figure}[bp]
\centering
\includegraphics[width=9cm]{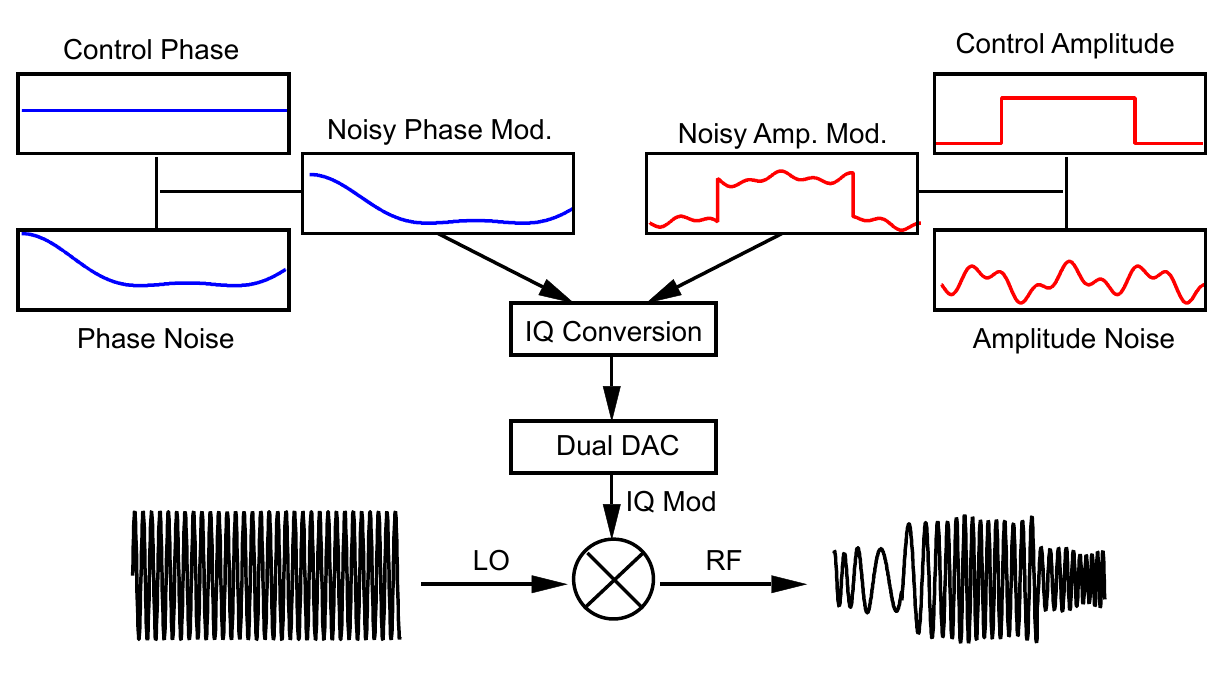}\\
\caption{(color online) Schematic representation of process flow involved in experimental noise engineering.  Independent waveforms for phase and amplitude are determined numerically, shifted to the $IQ$ basis and sent to the Dual DAC used for \emph{IQ} modulation in our vector signal generator. 
\label{Fig:IQ}}
\end{figure}

To model a desired control field in the presence of noise we generate a microwave field of the form set out in Eqs. \ref{MicrowaveField}-\ref{MicrowavePhase}. 
%\begin{align}
%\label{MicrowaveField}\MicroField(t) &= \MicroAmp(t)\cos(\MicroCarrierFreq t+\phi(t))\\
%\label{MicrowaveAmplitude}\MicroAmp(t) &= \MicroAmpControl(t)+\MicroAmpNoise(t)\\
%\label{MicrowavePhase}\phi(t) &= \phiControl(t)+\phiNoise(t).
%\end{align}
%\begin{align}
%\Omega(t)=\Omega_{C}(t)+\Omega_{N}(t)\\
% \phi(t)=\phi_{C}(t)+\phi_{N}(t) \nonumber
% \end{align}
%\noindent  where the subscripts $C$ and $N$ denote the desired control and added noise respectively. The first line is formally equivalent to Eq.~\ref{AmpNoise}, but is written using notation more convenient from the experimental perspective.  For instance, a noise-free $\pi$ pulse about the $x$ axis would have $\Omega_{N}=\phi_{N}=0$, and $\phi_{C}=0$, $\Omega_{C}(t)=\Omega$ for $t\in[0,\tau_{\pi}]$, with $\tau_{\pi}$.
In order to engineer the bath in our experimental system we begin with a desired noise power spectral density in either the amplitude or detuning quadrature (or both), assuming they are statistically independent.  From this power spectrum, defined by the noise strength $\alpha$, the exponent of the power-law scaling $p$, the comb spacing $\omega_{0}$, and the high-frequency cutoff $\omega_{c}\geq J\omega_{0}$, we numerically generate time-domain vectors for $\Omega_{N}(t)$ and $\phi_{N}(t)$ using the relationships appearing in Sec.~\ref{Sec:Theory}, and randomly selecting the phase $\psi_{j}$ for each comb tooth in the Fourier decomposition.  Thus we may independently generate our control and noise modulation signals for the carrier amplitude and phase (see Fig.~\ref{Fig:IQ}).

Independent and arbitrary control over these properties of the carrier may be achieved using $IQ$ modulation ~\cite{BarryBook2004}.  $I(t)$ and $Q(t)$ are simply a polar-to-cartesian coordinate transform of the familiar amplitude $\Omega(t)$ and phase $\phi(t)$ components of a modulated signal $S(t)=\Omega(t)\sin\left(\MicroCarrierFreq t+\phi(t)\right)$ as 

\begin{align}
&S(t)=I(t)\sin(\MicroCarrierFreq t)-Q(t)\sin(\MicroCarrierFreq t-\pi/2)\\
&I(t)=\Omega(t)\cos(\phi(t)),\;\;\;
Q(t)=\Omega(t)\sin(\phi(t)).
\end{align}

\noindent The numerically generated noise and control modulation patterns are thus converted to the $IQ$ basis and applied as a modulation pattern in time.  While our method typically relies on a Fourier decomposition for the generation of the $IQ$ modulation patterns, arbitrary time-domain noise may be engineered, such as the influence of random telegraph noise in carrier frequency.  

As mentioned above, our carrier frequency for the clock transition in $^{171}$Yb$^{+}$ is 12.6 GHz, produced by a vector signal generator.  The key feature of this unit is a digitally programmable baseband generator producing the modulation envelopes for $I$ and $Q$.  The functions are defined sample-wise with 16-bit resolution in order to approximate a continuous function.  In our system, care must be taken to ensure that discontinuities in the waveforms are avoided as the baseband generator employes an interpolation algorithm that can produce ringing in the \emph{applied} modulation.

%\begin{figure}[hbtp]
%\centering
%\includegraphics[width=15cm]{XXX.pdf}\\
%\caption{(color online) Measured phase noise for various engineered bath spectra.

%\label{Fig:ExptSchem}}
%\end{figure*}

%we demonstrate IQ modulated carrier signals for both amplitude and detuning noise as measured using a vector signal analyser. To produce each subfigure we inject noise into 0.5 s of constant amplitude and constant phase microwaves at 12.6 GHz and observe the phase noise spectrum using a vector signal analyzer. The noise is synthesized as prescribed in sections A and B. We show the phase noise spectra for $1/f^{2}$ amplitude noise, white amplitude noise, 1/f dephasing noise, and white dephasing noise in figures XXX, XXX, XXX, and XXX respectively. The horizontal axis gives the deviation from the carrier frequency. The vertical axis is in  Table XXX gives the corresponding parameters used in the constructions of the noise comb spectra, as well as the respective spectral fall off power as determine by the fits.

%				---------------------------------
				%subSECTION: Direct characterization of microwave carrier
%				---------------------------------

\subsection{Direct characterization of the microwave carrier}
In order to quantitatively verify the noise engineering process we begin by measuring the resultant phase noise on a 12.6 GHz carrier in the presence of bath engineering.  Interpreting such data requires a brief quantitative analysis of the effect of amplitude and phase modulation as represented in the Fourier domain.  In the case of amplitude modulation, a signal consisting of a single-frequency amplitude modulated (AM) carrier can be expressed as

\begin{align}
S(t)&=\Big[A_{\mu}+A_{m}\sin(\omega_{m}t)\Big]\sin(\omega_{\mu}t)\\
&=A_{\mu}\sin(\omega_{\mu}t)+\frac{A_{m}}{2}\cos(\delta_{-}t)\nonumber-\frac{A_{m}}{2}\cos(\delta_{+}t)\\
\end{align}

\noindent where $A_{\mu}$ and $A_{m}$ are the amplitudes of the carrier and modulating sinusoid and $\omega_{\mu}$ and $\omega_{m}$ are their frequencies, respectively. In the second line above we see the signal is represented as a weighted sum of the carrier frequencies as well as two symmetric sideband frequencies $\delta_{\pm}=\omega_{\mu}\pm\omega_{m}$ from the carrier.  Referring back to Eq.~\ref{GeneralizedAmplitudeNoise}, each comb tooth gives rise to a pair of sidebands with amplitude (power) proportional to $\alpha F(j)$ ($\alpha^{2}F(j)^{2}$). In this case the power-law scalings of the comb teeth ($p$) and the measured phase-noise power spectrum are identical.

The case of dephasing noise is slightly more complicated due to the effect of frequency or phase modulation as represented in the Fourier domain. Single-frequency phase modulation with amplitude $\Phi_{m}$ at frequency $\omega_{m}$ gives a signal 
\begin{align}
S(t)&=A_{\mu}\sin\Big(\omega_{\mu}t+\Phi_{m}\sin(\omega_{m}t)\Big)\\
&=A_{\mu}\sum\limits_{n=-\infty}^{\infty} J_{n}(\Phi_{m})\sin((\omega_{\mu}+n\omega_{m})t)\nonumber\\ 
&\approx A_{\mu}\sin(\omega_{\mu}t)+\frac{A_{\mu}\Phi_{m}}{2}\sin(\delta_{+}t)-\frac{A_{\mu}\Phi_{m}}{2}\sin(\delta_{-}t).\nonumber
\end{align}
\noindent Such modulation produces an infinite comb of frequencies, centred around the carrier, spaced by $\omega_{m}$, and weighted by Bessel functions ~\cite{ChowningJAES1973}.  In the last line above we have assumed small modulation depth allowing the infinite comb to be truncated beyond first order.  As a result of this expression the relationship between the power-law of comb teeth in Eq.~\ref{PSDCombMain} and the expected form of the phase noise is $p\leftrightarrow(p-2)$.

Phase-noise power spectra are presented in Fig.~\ref{Fig:NoiseData}a-d using units of dBc/Hz as a function of offset from the carrier, for different forms of bath engineering.  These data provide a measure of the total power at a particular offset referenced to the carrier in a one-Hertz integration bandwidth. In all cases we observe a strong increase in the measured phase noise over the (unmodulated) carrier noise floor up to a cutoff frequency corresponding to the programmed $\omega_{c}$. The form of decay in the phase-noise power law is well described using the expressions above, as indicated by guides to the eye superimposed on the measured data.  Agreement is good for both amplitude and detuning noise.   We also observe that as the noise strength (\emph{i.e.} modulation depth) increases for dephasing noise, the first order approximation above fails and the cutoff frequency is no longer clearly visible in these plots due to the infinite comb of sidebands.

\begin{figure}[htbp]
\centering
\includegraphics[width=7.5cm]{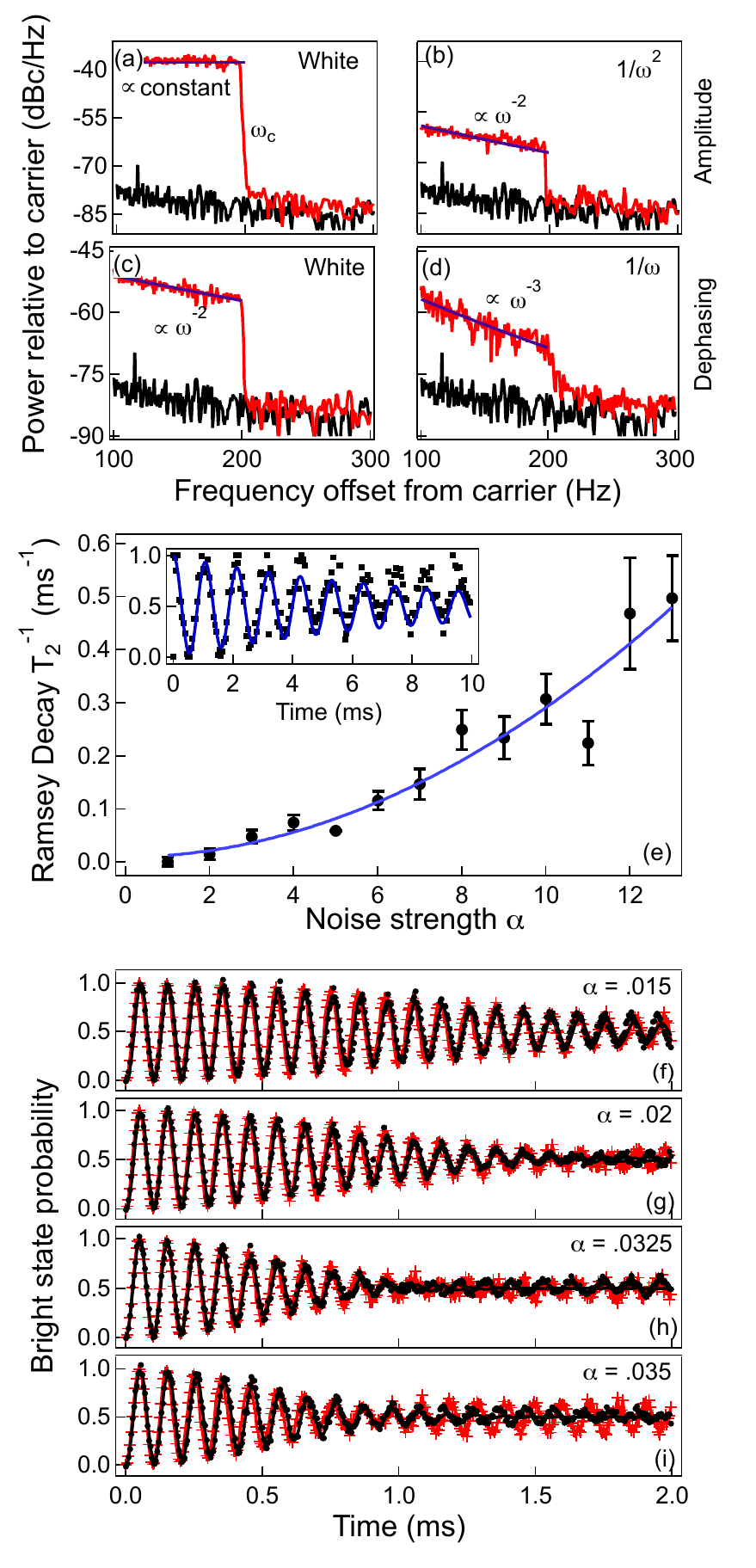}\\
\caption{(color online) Experimental validation of noise engineering.   a-d) Phase noise spectrum of carrier with engineered noise measured using a vector signal analyzer.  Black data represent the unmodulated noise floor at 12.6 GHz, while the red represent data with engineered bath.  Blue overlays are guides to the eye demonstrating the predicted phase noise scaling associated with particular power spectra.  e) Scaling of measured $T_{2}$ as a function of noise strength, $\alpha$, for white dephasing noise, $F(j)\propto j^{-1}$.  Error bars are determined from the fitting procedure employed in analyzing Ramsey data and blue line represents a quadratic fit, in line with expectations (see text).  Inset) Representative Ramsey data with overlaid exponential decay.  Data taken with fixed carrier detuning $\sim$1 kHz in order to show Ramsey fringes. f-i) Rabi oscillations in the presence of amplitude noise with a white power spectrum measured in experiment (black markers) and calculated via numerical integration of the Schrodinger equation (red markers) for different noise strengths.   Line traces show Gaussian decay fits.\label{Fig:NoiseData}}
\end{figure}

%\begin{align}
%S(t)&=A_{c}\sin(\omega_{c}t)\cos(\Phi_{m}\sin(\omega_{m}t))\nonumber\\
%&~~~+A_{c}\cos(\omega_{c}t)\sin(\Phi_{m}\sin(\omega_{m}t))\\
%&\approx A_{c}\sin(\omega_{c}t)+A_{c}\cos(\omega_{c}t)\Phi_{m}\sin(\omega_{m}t)\\
%&=A_{c}\sin(\omega_{c}t)+\frac{A_{c}\Phi_{m}}{2}\sin((\omega_{c}+\omega_{m})t)\nonumber\\
%&~~~~~~~~~~~~~~~~~~-\frac{A_{c}\Phi_{m}}{2}\sin((\omega_{c}-\omega_{m})t)
%\end{align}

%\begin{table}[bp]\label{}
%  \centering
%\begin{tabular}{c|c|c||c|c|}
%\cline{2-5}
% & \multicolumn{2}{|c||}{Dephasing}  &\multicolumn{2}{c|}{Amplitude}\\
%\cline{2-5}
%&  $1/f$ &White &  $1/f^{2}$ &White  \\
%\hline
%\multicolumn{1}{|c|}{\parbox{2 cm}{~\\$p$\\~}} &\parbox{1 cm}{$-1$} &\parbox{1 cm}{$0$}   &\parbox{1 cm}{$-2$} & \parbox{1 cm}{$0$}    \\ 
%\multicolumn{1}{|c|}{\parbox{2 cm}{~\\$\alpha$\\~}} &\parbox{1 cm}{$5$} &\parbox{1 cm}{$1$}   &\parbox{1 cm}{$0.25$} &\parbox{1 cm}{$0.01$}  \\ 
%\multicolumn{1}{|c|}{\parbox{2 cm}{~\\Expected\\decay power}} &$-3$ & $-2$   &$-2$ & $0$  \\ 
%\multicolumn{1}{|c|}{\parbox{2 cm}{~\\Measured\\decay power}} &$-2.7\pm0.4$ & $-1.7\pm0.2$   &$-2.0\pm0.2$ & $0.2\pm0.2$  \\
%\hline
%\end{tabular}
%\caption{Experimental parameters for noise spectrum measurements.}\label{NoisePowerLaws}
%\end{table}

%				---------------------------------
				%subSECTION: Qubits in a noisy bath
%				---------------------------------

\subsection{Qubits in a noisy bath}
We quantitatively demonstrate our bath engineering techniques by studying the effect of engineered dephasing noise on the free-evolution of our trapped-ion qubits.   We produce time-domain dephasing noise using a discrete comb with fundamental frequency $\omega_{0}=2\pi\times4$ Hz, and a cutoff $\omega_{c}=2\pi\times 3$ kHz.  Selecting $F(j)=j^{-1}$, yields a white dephasing noise power spectral density.

In our experiments we perform Ramsey spectroscopy in the presence of engineered noise.  We observe the decay time constant of fringe visibility is significantly reduced in the presence of the engineered bath, as expected.  A representative Ramsey experiment is presented in the inset of Fig.~\ref{Fig:NoiseData}e showing the fringe visibility decays over a timescale of order milliseconds in the presence of the engineered bath. Since Ramsey spectroscopy measures the relative coherence between two oscillators (the LO and the qubit), engineered dephasing in the LO results in net decoherence even during free-evolution periods. 

The scaling of the measured coherence time with noise strength is a key validation of our bath engineering techniques.  We write the ensemble averaged coherence for free evolution in the presence of dephasing noise as $W(t)=\exp[-\chi(\tau)]$ ~\cite{BiercukPRA2009,UhrigNJP2008} where $\chi(\tau)=\frac{2}{\pi}\int_{0}^{\infty}\frac{S_{\beta}(\omega)}{\omega^{2}}\sin^{2}(\omega\tau/2)d\omega$ ~\cite{BiercukJPB2011}. Incorporating the relevant form of $S_{\beta}(\omega)$ gives

\begin{align}
\chi(\tau)&=2\alpha^{2}\omega_{0}^{2}\sum_{j=1}^{750}\frac{\sin^{2}(j\omega_{0}\tau/2)}{(j\omega_{0})^{2}}
\end{align}
\noindent where the upper limit on the sum is determined from the specifics of the comb tooth spacing and noise cutoff frequency.  The expected decay envelope resulting from the integral above and for our value of $\omega_{0}$ is a simple exponential, although slight modification yields more complex functional forms (see Appendix \ref{ApdxSec: Chi(tau)}).  As a result we expect a quadratic scaling of the decay time constant with $\alpha$.

We study the scaling of the $1/e$ coherence time, $T_{2}$, as a function of the noise strength, parameterized by $\alpha$.  Ramsey fringes are recorded for each value of $\alpha$ and a fit to a sinusoid with a simple exponential decay envelope is performed in order to extract $T_{2}$.  These data are plotted as the decay \emph{rate}, $T_2^{-1}$, of the qubit coherence as a function of $\alpha$ in Fig.~\ref{Fig:NoiseData}e.  The decay rate is observed to scale as $T_2^{-1}\propto\alpha^2$, as expected, with the overall decay timescales determined by the specifics of the noise power spectral density.  

Similarly we study driven evolution such as that presented in Fig. 2c in the presence of engineered amplitude noise (\label{Fig:NoiseData}f-i). Increasing the strength of the noise results in a Gaussian decay envelope for the Rabi flopping data with decreasing decay constants.  The evolution of the measured decay constant does not take a convenient analytic form  with noise strength and so we perform numerical integration of the Schrodinger equation considering a driving field with the same noise characteristics.  The data show good agreement with the numerical calculations qualitatively and quantitatively.

Other experiments incorporating dephasing noise during driven evolution or and universal noise also reveal behavior in quantitative agreement with the formulation provided above, but form the subject of separate manuscripts.  Overall these measurements validate the efficacy of our approach in generating a quantitatively useful noise bath in a real quantum system.  

%__________________________________________________________________________________________

%				SECTION: CONCLUSION
%__________________________________________________________________________________________

\section{Conclusion}\label{Sec:Conclusion}

In this manuscript we have presented a detailed technical prescription for the quantitative engineering of unitary baths for studies of quantum control.  We produce a discrete comb of noise frequencies possessing an overall scaling chosen to reproduce a noise power spectrum of interest in either the dephasing or amplitude damping quadrature.  We show how this technique lends itself to simple numerical construction of complex time-dependent noise processes using common $IQ$-modulated carriers for single-qubit control.  We validate our technique through both examination of the modulation form on a vector signal analyser and through application of engineered dephasing noise to the free evolution of trapped $^{171}$Yb$^{+}$ ions.  Our measurements demonstrate that the coherent lifetime of the qubits probed by a 12.6 GHz carrier incorporating engineered noise scales as expected based on a simple physical model.

The technique we present is applicable to a wide variety of experimental systems employing carrier signals in the RF or microwave.  It is particularly useful in trying to understand the spectral sensitivity of various quantum control protocols such as dynamical decoupling and dynamically corrected gates.  For instance, our group has utilized this technique to quantitatively validate the error-suppressing properties of novel classes of modulated error-suppressing gates, as will be described in future manuscripts.  The incorporation of engineered noise in such experiments is vital to help elucidate the bounds and performance scaling of such protocols in regimes where the measured errors (the signal of interest) are in general not sufficiently large to exceed intrinsic state-preparation and measurement errors.  It is also possible to combine this kind of unitary noise engineering with dissipation~\cite{ViolaPointer, Irfan_PRL2012, BlattDissipative, WinelandDissipative}, for instance, through leakage of off-resonant lasers or otherwise inducing spontaneous emission properties, or to expand the general technique to multi-level manifolds ~\cite{JessenPRL2013}.  We hope that our general technique will prove useful to the quantum control and quantum information communities as they push towards ultra-high fidelity gate operations.

\begin{acknowledgments}
This work partially supported by the US Army Research Office under Contract Number  W911NF-11-1-0068, and the Australian Research Council Centre of Excellence for Engineered Quantum Systems CE110001013, the Office of the Director of National Intelligence (ODNI), Intelligence Advanced Research Projects Activity (IARPA), through the Army Research Office, and the Lockheed Martin Corporation. All statements of fact, opinion or conclusions contained herein are those of the authors and should not be construed as representing the official views or policies of IARPA, the ODNI, or the U.S. Government.
\end{acknowledgments}

%__________________________________________________________________________________________

%							     BIBLIOGRAPHY
%__________________________________________________________________________________________
\newpage

%BIBLIOGRAPHY

\bibliographystyle{apsrev4-1}
\bibliography{NoiseEngineeringBib}

%						__________________________________

%							     APPENDICES
%						__________________________________
\newpage
\appendix
\begin{widetext}

\section{Derivation of the System Hamiltonian\label{Sec:AppendixA}}
The system considered in this paper consists of a qubit with transition (angular) frequency $\Qsplitting$, driven by a local oscillator with magnetic field component aligned with $\sigz$ taking the form 
\begin{align}
\label{MicrowaveFieldA}\MicroField(t) &= \MicroAmp(t)\cos(\MicroCarrierFreq t+\phi(t))\hat{\mathbf{z}}\\
\label{MicrowaveAmplitudeA}\MicroAmp(t) &= \MicroAmpControl(t)+\MicroAmpNoise(t)\\
\label{MicrowavePhaseA}\phi(t) &= \phiControl(t)+\phiNoise(t)
\end{align}
with $\MicroCarrierFreq$ the carrier frequency.  In this formulation the time-dependence of the phase $\phi(t)$ and amplitude $\MicroAmp(t)$ has been formally partitioned into components denoted by the subscripts $C$ and $N$, capturing the desired control and noisy interactions respectively.  Using the dipole approximation, the system Hamiltonian ($\hbar=1$) in the laboratory frame is
\begin{align}
&H_S = \frac{\Qsplitting}{2}\sigz + \Omega(t)\cos(\omega_\mu t + \phi(t))\sigx.
\end{align}
The first term corresponds to the Hamiltonian of the qubit under free evolution, and the second term corresponds to the qubit-field interaction. Moving to the interaction picture co-rotating with the carrier frequency and making the rotating-wave approximation, the dynamics are described by the Hamiltonian
\begin{align}\label{CarrierPictureHamiltonianA}
H_I^{(\omega_\mu)}=\Big(\frac{\Delta}{2}\Big)\sigz +
\frac{1}{4}\Omega(t)
\Big\{
e^{-i\phi(t)}\sigp+
e^{i\phi(t)}\sigm
\Big\}
\end{align}
where $\Delta = \Qsplitting-\omega_\mu$ is the carrier detuning from the transition frequency, and we define the usual operators $\hat{\sigma}_{\pm} = \sigx\pm i\sigy$.
%\footnote{Setting $\phi_N(t) = 0$ and $\Delta = 0$, Eq. \ref{CarrierPictureHamiltonian} reduces to $H_I^{(\omega_\mu)}=
%\frac{1}{2}\Omega_R \big[
%\cos(\phi_0)\sigx+\sin(\phi_0)\sigy\big]
%$, corresponding to coherent Rabi flopping about the axis $\vec{\mathbf{r}} = (\cos(\phi_0),\sin(\phi_0),0)$ with Rabi rate characterized by $\Omega_R$.}.
Setting $\Delta=\phi_N(t)=0$ Eq. \ref{CarrierPictureHamiltonianA} reduces to $H_I^{(\omega_\mu)}=
\frac{1}{2}\Omega(t)\big[
\cos(\phi_C(t))\sigx+\sin(\phi_C(t))\sigy\big]
$, so the resonant carrier field drives coherent Rabi flopping between the qubit states $\ket{1}$ and $\ket{0}$. The instantaneous Rabi rate in this case is proportional to $\MicroAmp(t)$ and rotations, generated by the spin operator $\hat{\sigma}_{\phi_C(t)}$, are driven about the axis $\vec{r}=\left(\cos\phi_C(t), \sin\phi_C(t),0\right)$ in the $xy$-plane of the Bloch sphere.  For instance, a noise-free $\pi$ pulse about the $x$-axis would have $\Omega_{N}, \phi_{N}, \phi_C=0$, and $\Omega_{C}(t)=\Omega$ for $t\in[0,\tau_{\pi}]$, with $\tau_{\pi} = \pi/\Omega$.

The noise component $\phi_N(t)$ of the engineered phase $\phi(t)$, in the exponentials of Eq. \ref{CarrierPictureHamiltonianA}, may be mapped to rotations about $\sigz$ if we move to a second interaction picture defined by 
\begin{align}
H_I^{(\omega_\mu,\dot{\phi}_N)} := U_{\dot{\phi}_N}^\dagger H_I^{(\omega_\mu)} U_{\dot{\phi}_N} - \HphiN,
\end{align}
where $U_{\dot{\phi}_N}(t) = \exp[-i\frac{\phi_N(t)}{2}\sigz]$ is the evolution operator under the the engineered dephasing Hamiltonian $\HphiN := \frac{1}{2}\dot{\phi}_N(t)\sigz$. Setting the carrier detuning $\Delta = 0$ 
%and defining the engineered detuning noise field $\Bd(t)\equiv-\frac{1}{2}\dot{\phi}_N(t)$, 
it is then straightforeward to show
\begin{align}\label{PhiNnteractionHamiltonianA}
H_I^{(\omega_\mu,\dot{\phi}_N)}= -\frac{\dot{\phi}_N(t)}{2}\sigz +
\frac{1}{2}\Omega(t)
\Big\{
\cos[\phi_C(t)]\sigx+
\sin[\phi_C(t)]\sigy
\Big\}.
\end{align}

%__________________________________________________________________________________________

%				  SECTION: Derivation of Noise Power Spectral Densities
%__________________________________________________________________________________________
\section{Derivation of Noise Power Spectral Densities}\label{ApdxSec: PSDs}
Let $h(t)$ be any time-dependent function. We use non-unitary angular frequency notation consistent with the usage in Ref.~\cite{GreenNJP2013} to define a Fourier transform pair 
\begin{align}\label{FourierTransformPair1}
H(\omega) &= \FT\big[h(t)\big] = \int_{-\infty}^\infty h(t)e^{-i\omega t} dt\\
\label{FourierTransformPair2}h(t) &= \FT^{-1}\big[H(\omega)\big] = \frac{1}{2\pi}\int_{-\infty}^\infty H(\omega) e^{i\omega t} d\omega.
\end{align}
In this case, a time-domain signal $\beta(t)$ is related to its PSD $S_\beta(\omega)$ by the Wiener-Khintchine Theorem~\cite{MillerBook2012} which takes the form
\begin{align}
\langle\beta(t_1)\beta(t_2)\rangle&= \frac{1}{2\pi}\int_{-\infty}^\infty d\omega S_\beta(\omega) e^{i\omega(t_2-t_1)}.
\end{align}
In this paper we assume all noise processes are \emph{wide-sense stationary} in which case the two-point corrleation function $\langle\beta(t_1)\beta(t_2)\rangle$ depends only on the \emph{relative difference} $\tau = t_2-t_1$ between $t_1$ and $t_2$ and reduces to the auto-correlation function $C_\beta(\tau): = \langle\beta(t_1)\beta(t_1+\tau)\rangle$. The angle brackets now denote averaging over all times $t_1$ with respect to which the relative lag $\tau$ is defined. Consequently  $C_\beta(\tau)= \frac{1}{2\pi}\int_{-\infty}^\infty d\omega S_\beta(\omega) e^{i\omega\tau}$, or using Eqs. \ref{FourierTransformPair1} and \ref{FourierTransformPair2}, 
\begin{align}\label{PSDfromAutoCorrelation}
S_\beta(\omega) = \FT\big[C_\beta(\tau)\big]= \int_{-\infty}^\infty \langle\beta(t)\beta(t+\tau)\rangle e^{-i\omega \tau} d\tau.
\end{align}

%				---------------------------------
				%subSECTION: Dephasing Noise
%				---------------------------------
\subsection{Dephasing Noise PSD}\label{Appendix:DephasingPSD}

We require an expression for the auto-correlation function of $\Bd(t)$, the dephasing noise field, in order to derive the correpsonding PSD $\Sd(\omega)$ given by Eq. \ref{PSDfromAutoCorrelation}. Using Eq. \ref{DetuningNoise} it is straightforward to show
\begin{align}
\Bd(t+\tau)\Bd(t)=\frac{\alpha^2\omega_0^2}{4}\sum_{j,j'=1}^Jjj'F(j)F(j')
\Big[e^{i\omega_j \tau}e^{i(\omega_j +\omega_{j'}) t}e^{i(\psi_j+\psi_{j'})}
+e^{i\omega_j \tau}e^{i(\omega_j -\omega_{j'})t}e^{i(\psi_j-\psi_{j'})}+c.c\Big].
\end{align}
Or, averaging over $t$,
\begin{align*}
\langle\Bd(t+\tau)\Bd(t)\rangle_t=\frac{\alpha^2\omega_0^2}{4}\sum_{j,j'=1}^Jjj'F(j)F(j')
&\Big[e^{i\omega_j \tau}e^{i(\psi_j+\psi_{j'})}\langle e^{i(\omega_j +\omega_{j'}) t}\rangle_t+e^{i\omega_j \tau}e^{i(\psi_j-\psi_{j'})}\langle e^{i(\omega_j -\omega_{j'})t}\rangle_t\nonumber\\
&+e^{-i\omega_j \tau}e^{-i(\psi_j-\psi_{j'})}\langle e^{-i(\omega_j-\omega_{j'})t}\rangle_t+e^{-i\omega_j \tau}e^{-i(\psi_j+i\psi_{j'})}\langle e^{-i(\omega_j+\omega_{j'})t}\rangle_t
\Big].
\end{align*}
Since $\omega_j,\omega_{j'}$ are always positive we know $\omega_j+\omega_{j'}$ is always positive.  Consequently the term $e^{\pm i(\omega_j+\omega_{j'})t}$ is always an oscillating term with nonzero frequency $\pm (\omega_j+\omega_{j'})$, and average over $t$ yields
\begin{align}
\langle e^{\pm i(\omega_j+\omega_{j'})t}\rangle_t = 0.
\end{align}
Similarly, when $\pm(\omega_j-\omega_{j'})$ is nonzero (i.e. when $\omega_j\ne\omega_{j'}\iff j\ne j'$), we have  
\begin{align}
\langle e^{\pm i(\omega_j-\omega_{j'})t}\rangle_t = 0,\hspace{1cm}j\ne j'.
\end{align}
However, when $\omega_j=\omega_{j'}$ (which occurs when $j=j'$)
\begin{align}
\langle e^{\pm i(\omega_j-\omega_{j'})t}\rangle_t  = 1,\hspace{1cm}j= j'
\end{align}
%\begin{align}
%\langle e^{\pm i(\omega_j-\omega_{j'})t}\rangle_t = \langle e^{\pm i0}\rangle_t = \langle \cos(0)\rangle_t = \langle 1\rangle_t  = 1,\hspace{1cm}j= j'
%\end{align}
or more concisely
\begin{align}
\langle e^{\pm i(\omega_j-\omega_{j'})t}\rangle_t = \delta_{jj'}
\end{align}
where $\delta_{ij}$ is the Kronecker delta. Thus the auto-correlation function for $\Bd(t)$ takes the form
\begin{align}
\langle\Bd(t+\tau)\Bd(t)\rangle_t&=\frac{\alpha^2\omega_0^2}{4}\sum_{j,j'=1}^Jjj'F(j)F(j')\Big[e^{i\omega_j \tau}e^{i(\psi_j-\psi_{j'})} \delta_{jj'}+e^{-i\omega_j \tau}e^{-i(\psi_j-\psi_{j'})}\delta_{jj'}\Big]\\
&=\frac{\alpha^2\omega_0^2}{4}\sum_{j=1}^Jj^2(F(j))^2\Big[e^{i\omega_j \tau}+e^{-i\omega_j \tau}\Big]\\
&=\frac{\alpha^2\omega_0^2}{2}\sum_{j=1}^J(jF(j))^2\cos(\omega_j \tau).
\end{align}
%\subsection{Computing $\Sd(\omega)$}
Substituting this into Eq. \ref{PSDfromAutoCorrelation} yields
\begin{align}
\Sd(\omega) &= \int_{-\infty}^\infty \langle\Bd(t)\Bd(t+\tau)\rangle_te^{-i\omega \tau} d\tau\\
&= \int_{-\infty}^\infty \Big[\frac{\alpha^2\omega_0^2}{2}\sum_{j=1}^J(jF(j))^2\cos(\omega_j \tau)\Big]e^{-i\omega \tau} d\tau\\
%&= \frac{\alpha^2\omega_0^2}{2}\sum_{j=1}^J(jF(j))^2\int_{-\infty}^\infty \cos(\omega_j \tau) e^{-i\omega \tau} d\tau\\
&= \frac{\alpha^2\omega_0^2}{2}\sum_{j=1}^J(jF(j))^2\FT\big[\cos(\omega_j \tau)\big] 
\end{align}
Using the result from Fourier analysis that 
\begin{align}\label{FourierTrigIdentity}
\FT\big[\cos(\omega' \tau)\big] &= \int_{-\infty}^\infty \cos(\omega'\tau)e^{-i\omega \tau} d\tau = \pi(\delta(\omega-\omega')+\delta(\omega+\omega'))
\end{align}
we therefore obtain our result
\begin{align}
\Sd(\omega) &= \frac{\alpha^2\omega_0^2}{2}\sum_{j=1}^J(jF(j))^2\Big[
\pi(\delta(\omega-\omega_j)+\delta(\omega+\omega_j))
\Big].
\end{align}

%				---------------------------------
				%subSECTION: Amplitude Noise
%				---------------------------------

\subsection{Amplitude Noise PSD}

We require an expression for the auto-correlation function of $\Ba(t)$, the amplitude noise field, in order to derive the correpsonding PSD $\Sa(\omega)$ given by Eq. \ref{PSDfromAutoCorrelation}. Using Eq. \ref{GeneralizedAmplitudeNoise}, and following the same procedure used in the above section, it is straightforward to show
\begin{align}
\langle\Ba(t+\tau)\Ba(t)\rangle_t=\frac{\alpha^2}{2}\sum_{j=1}^J(F(j))^2\cos(\omega_j \tau).
\end{align}
%\subsection{Computing $\Sa(\omega)$}
Substituting this into Eq. \ref{PSDfromAutoCorrelation} and using Eq. \ref{FourierTrigIdentity} we therefore obtain the result
\begin{align}
\Sa(\omega) &= \frac{\alpha^2}{2}\sum_{j=1}^J(F(j))^2\FT\big[\cos(\omega_j \tau)\big]\\ 
&= \frac{\pi\alpha^2}{2}\sum_{j=1}^J(F(j))^2\Big[\delta(\omega-\omega_j)+\delta(\omega+\omega_j)\Big].
\end{align}

%__________________________________________________________________________________________

%				  SECTION: DEPENDENCE OF CHI^2 ON tau
%__________________________________________________________________________________________

\section{Dependence of $\chi$ on $\tau$ for FID}\label{ApdxSec: Chi(tau)}

In the case of a pure white noise detuning PSD it is relatively simple to calculate an exact form for $\chi(\tau)$. Starting with $\chi(\tau)=\frac{2}{\pi}\int_{0}^{\infty}\frac{S_{\beta}(\omega)}{\omega^{2}}\sin^{2}(\omega\tau/2)d\omega$ and incorporating $S_{\beta}(\omega)=\alpha^{2}$ gives

\begin{align}
\chi(\tau)&=\frac{2\alpha^{2}}{\pi}\int_{0}^{\infty}\frac{\sin^{2}(\omega\tau/2)}{\omega^{2}}d\omega=\frac{\tau\alpha^{2}}{2}
\end{align}

\noindent giving $\chi$ a linear dependence on $\tau$, and hence producing a simple exponential decay in fidelity (fringe visibility).  By contrast, in the limit of weak low-frequency-dominated noise, use of the small angle approximation for the sinusoidal term results in a quadratic dependence, $\chi(\tau)\propto \alpha^{2}\tau^{2}$, yielding a Gaussian decay envelope in Ramsey fringe visibility.  

The choice of fit-function for the Ramsey decay therefore depends sensitively on the details of the noise model employed.  Our engineered white noise PSD consists of a finite set of comb teeth spaced by a finite frequency interval designed to approximate a continuous white noise PSD with a finite cutoff frequency. In section III.D. we defined a noise power spectrum producing a coherence integral

\begin{align}
\chi(\tau)&=2\alpha^{2}\omega_{0}^{2}\sum_{j=1}^{750}\frac{\sin^{2}(j\omega_{0}\tau/2)}{(j\omega_{0})^{2}}
\end{align}

\noindent where $\omega_{0}=4$ Hz. In such circumstances we rely on numerical calculations to determine the behavior $\chi(\tau)$ over the time interval relevant to the Ramsey spectroscopy experiments in Fig.~\ref{Fig:TauDependence}a.   For our choice of $\omega_{0}$ we find good approximation of the integral to a linear function of $\tau$, suggesting the use of a simple exponential fit to Ramsey decay in FID experiments with engineered noise.  However, we observe that modifying the fundamental frequency of the power spectrum changes the dependence of $\chi(\tau)$ within the same evolution time interval, requiring careful attention to the noise model in use when performing quantitative studies of bath engineering.
\begin{figure}[htbp]
\centering
\includegraphics[width=12cm]{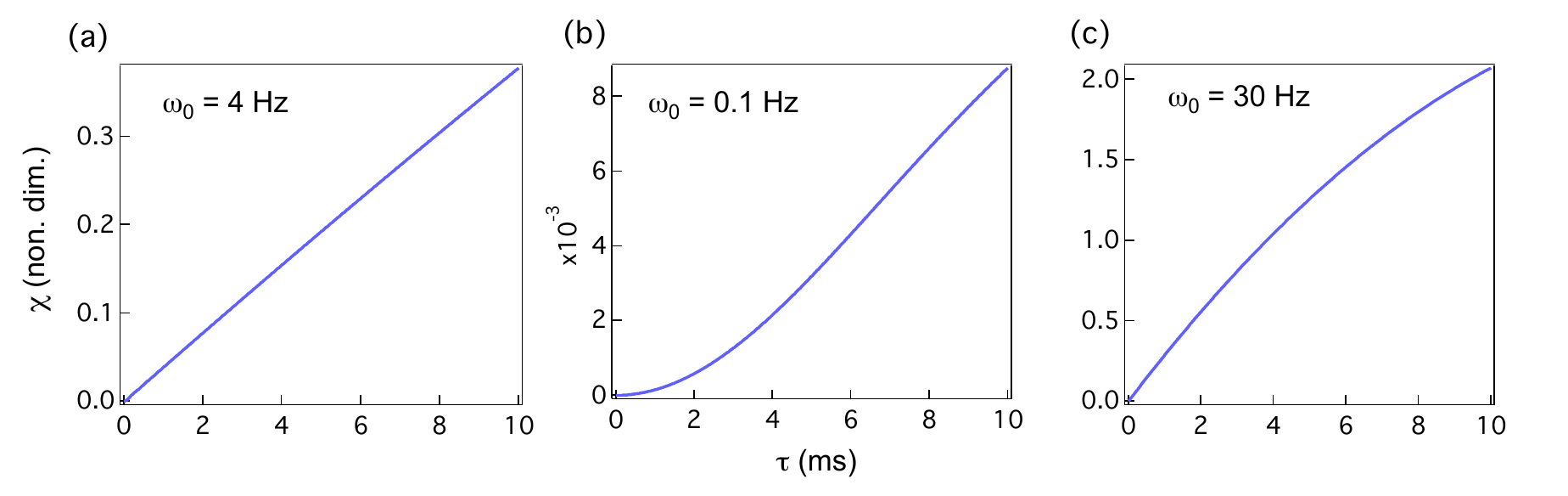}\\
\caption{Dependence of $\chi(\tau)$ on $\tau$ for various choices of $\omega_{0}$.  $\chi(\tau)$ is calculated for free evolution in the presence of engineered white detuning noise.  The $\tau$ axis runs over the time period used in the experiments described in the main text.  Without loss of generality we eliminate $\alpha$ as a free parameter by setting it equal to 1.
\label{Fig:TauDependence}}
\end{figure}

\end{widetext}
\end{document}